  \documentclass[final,5p,times,twocolumn]{elsarticle}

\usepackage{epsfig}

\usepackage{amssymb}
 \usepackage{amsthm}
 \usepackage{amsmath}
\usepackage{xcolor}  

\journal{Chinese Physics C}

\begin{document}

\begin{frontmatter}



\title{ISOMERIC RATIO OF THE $^{181}\rm{Ta}(\gamma,3n)^{178m,g}\rm{Ta}$ REACTION PRODUCTS \\
  AT ENERGY $E_{\rm{\gamma max}}$ UP TO 95 MeV }

\author{ O.S. Deiev} 

\author{I.S. Timchenko\corref{cor1}}
\ead{timchenko@kipt.kharkov.ua}
 \cortext[cor1]{Corresponding author}
\author{S.N. Olejnik,  \\ V.A. Kushnir, V.V. Mytrochenko, S.A. Perezhogin}
\address{National Science Center "Kharkiv Institute of Physics and Technology", \\
 1 Akademicheskaya St., 61108 Kharkiv, Ukraine}

\begin{abstract}
The photoneutron reaction $^{181}\rm{Ta}(\gamma,3n)^{178m,g}\rm{Ta}$ was investigated with the beam from the NSC KIPT electron linear accelerator LUE-40. The measurements were performed using the residual $\gamma$-activity method. The bremsstrahlung flux-averaged cross-sections $\langle{\sigma(E_{\rm{\gamma max}})}\rangle$, $\langle{\sigma(E_{\rm{\gamma max}})}\rangle_{\rm{m}}$, $\langle{\sigma(E_{\rm{\gamma max}})}\rangle_{\rm{g}}$ and the isomeric ratio of the reaction products $d(E_{\rm{\gamma max}})$ have been measured. Theoretical values of averaged cross-sections and isomeric ratio were calculated using partial cross-sections from the TALYS1.95 code for different level density models $LD$~1-6. The obtained experimental $d(E_{\rm{\gamma max}})$ agree with the literature data, but differ from the theoretical values in absolute magnitude and the behavior of the energy dependence. A comparison of the found averaged cross-sections with the calculated ones showed the best agreement for the case of the $LD$ 5 model.
\end{abstract}




\end{frontmatter}

\section{Introduction}
\label{Int}
Photonuclear reactions are accompanied by the emission of a nucleon or a group of nucleons from a compound nucleus. This leads to an excited state of the final nucleus, the discharge of excitation energy of which occurs in a time of $10^{-12}$--$10^{-17}$ sec. However, in some cases, at low energy of the excitation level and a high degree of forbidden transition, long-lived excited states of atomic nuclei occur. Such states are called isomeric or metastable, and their half-life can vary from nanoseconds to many years \cite{0}.

Nuclei with isomeric (m) and unstable ground states (g) are of particular interest, since they allow to study of the metastable state population of this nucleus relative to its ground state, i.e. obtain the isomeric ratio ($d(E)$ or $d(E_{\rm{\gamma max}})$) of the reaction products. This characteristic is defined as the ratio of the cross-sections for the formation of the reaction product in the metastable $\sigma_{\rm{m}}(E)$ and in the ground  $\sigma_{\rm{g}}(E)$ states: $d(E) = \sigma_{\rm{m}}(E)/\sigma_{\rm{g}}(E)$. The $d(E)$ values in the literature can be found as the ratio of the cross-sections for the formation of a product nucleus in the high-spin state (as a rule, this is a metastable state) $\sigma_{\rm{H}}(E)$ to the cross-section for the low-spin state $\sigma_{\rm{L}}(E)$: $d(E) = \sigma_{\rm{H}}(E)/\sigma_{\rm{L}}(E)$ \cite{1a,2a,3a}. In the case of research with the use of bremsstrahlung flux, this ratio $d(E)$ is expressed through the bremsstrahlung flux-averaged cross-sections or yields of the reactions under study \cite{3a,4a,5a}.

Data on isomeric ratios of reaction products make it possible to investigate issues related to nuclear reactions and nuclear structure, such as the spin dependence of the nuclear level density, angular momentum transfer, nucleon pairing, shell effects, refine the theory of gamma transitions, and test theoretical models of the nucleus \cite{1,2,3,4,5}. The study of isomeric ratios using photonuclear reactions has an advantage since the $\gamma$-quantum introduces a small angular momentum and does not change the nucleon composition of the compound nucleus.

The studies in the energy range above giant dipole resonance (GDR) and before the pion production threshold (30--145 MeV) are of interest, since the mechanism of the nuclear reaction changes here: from dominance of GDR to dominance of the quasideuteron mechanism \cite{15}. However, in this energy range, there is still a lack of experimental data on the cross-sections of multiparticle photonuclear reactions and isomeric ratios of the products of a nuclear reaction \cite{5b,6b}. This complicates the analysis of the mechanism of the nuclear reaction, as well as the systematization and analysis of the dependences of these quantities on various characteristics of the nucleus.

The experiments on photodisintegration of the $^{181}$Ta nucleus in the GDR region were carried out in many works \cite{6,7,8,9,10,11,12,13} using beams of quasi-monochromatic and bremsstrahlung photons. The partial cross-sections of photoneutron reactions at $^{181}$Ta were obtained by the method of direct registration of neutrons in the GDR region for  $(\gamma,\rm{n})$ and $ (\gamma,2\rm{n})$ in \cite{7}. In \cite{8}, studies of the partial cross-sections for the reactions  $(\gamma,\rm{n})$, $ (\gamma,2\rm{n})$ and  $(\gamma,3\rm{n})$, as well as for the reaction $(\gamma,4\rm{n})$ up to an energy of 36 MeV were carried out. New data for these reactions are presented in  \cite{16}. As for the study of the photodisintegration of tantalum at higher energies of bremsstrahlung $\gamma$-quanta, the results for  $E_{\rm{\gamma max}}$ = 55 MeV in the form of weighted average yields and at  $E_{\rm{\gamma max}}$ = 67.7 MeV in the form of a relative yield are presented in \cite{14} and \cite{15}, respectively. In \cite{19}, experimental values of the total bremsstrahlung flux-averaged cross-sections $\langle{\sigma(E_{\rm{\gamma max}})}\rangle$ were obtained for photoneutron reactions at $^{181}$Ta with emission of up to 8 neutrons at  $E_{\rm{\gamma max}}$ = 80--95 MeV and comparison was made with the calculations using the Talys1.9 code with default parameters.

Despite numerous experiments on the photodisintegration of  $^{181}$Ta \cite{16}, the issue of the isomeric ratio for the products of the reaction  $^{181}$Ta$(\gamma,3\rm{n})$ has not been sufficiently studied. Thus, the experimental values $d(E_{\rm{\gamma max}})$ for product nuclei $^{178\rm{m,g}}$Ta are presented in \cite{14,17,18}. From the data of work \cite{15} at  $E_{\rm{\gamma max}}$ = 67.7~MeV, it is possible to obtain an estimated value of $d(E_{\rm{\gamma max}})$ equal to 0.28 $\pm$ 0.08. In \cite{19}, experimental values of the isomeric ratio were found in the energy range  $E_{\rm{\gamma max}}$ = 80--95 MeV and the calculation was performed using the Talys1.9 code with default parameters. The comparison showed the excess of the calculated values over the experimental ones by two times. The data from \cite{14, 18}, measured at the same $\gamma$-quanta energy, differ significantly from each other. The results from \cite{17} and \cite{19} in the energy ranges of 22--32 and 80--95 MeV, respectively, show close values of the isomeric ratio.

In this work, we measured the bremsstrahlung flux-averaged cross-sections for the ground $\langle{\sigma(E_{\rm{\gamma max}})}\rangle_{\rm{g}}$ and isomeric $\langle{\sigma(E_{\rm{\gamma max}})}\rangle_{\rm{m}}$ states from the $^{181}\rm{Ta}(\gamma,3n)^{178m,g}\rm{Ta}$ reaction at the boundary energies of the bremsstrahlung spectra $E_{\rm{\gamma max}}$ = 35--80 MeV. Based on these data, the values of the total average cross-sections $\langle{\sigma(E_{\rm{\gamma max}})}\rangle$ and the isomeric ratio $d(E_{\rm{\gamma max}})$ of the reaction products were obtained. To calculate the theoretical values of the average cross-sections and $d(E_{\rm{\gamma max}})$, we used the partial cross-sections from the Talys1.95 code \cite{20} for different models of the level density $LD$~1–6. A comparison of the obtained experimental data is carried out, both with these calculations and with earlier literature data.

\section{Experimental procedure}
\label{Exp proced}

The experiments were performed using the method of measuring the residual $\gamma$-activity of the irradiated sample, which makes it possible to simultaneously obtain data on different channels of photonuclear reactions. This method is well known and described in several studies on multiparticle photonuclear reactions, for example, on the $^{93}$Nb nucleus \cite{21,22}.

\begin{figure}[h]
   	\center{\includegraphics[scale=0.3]{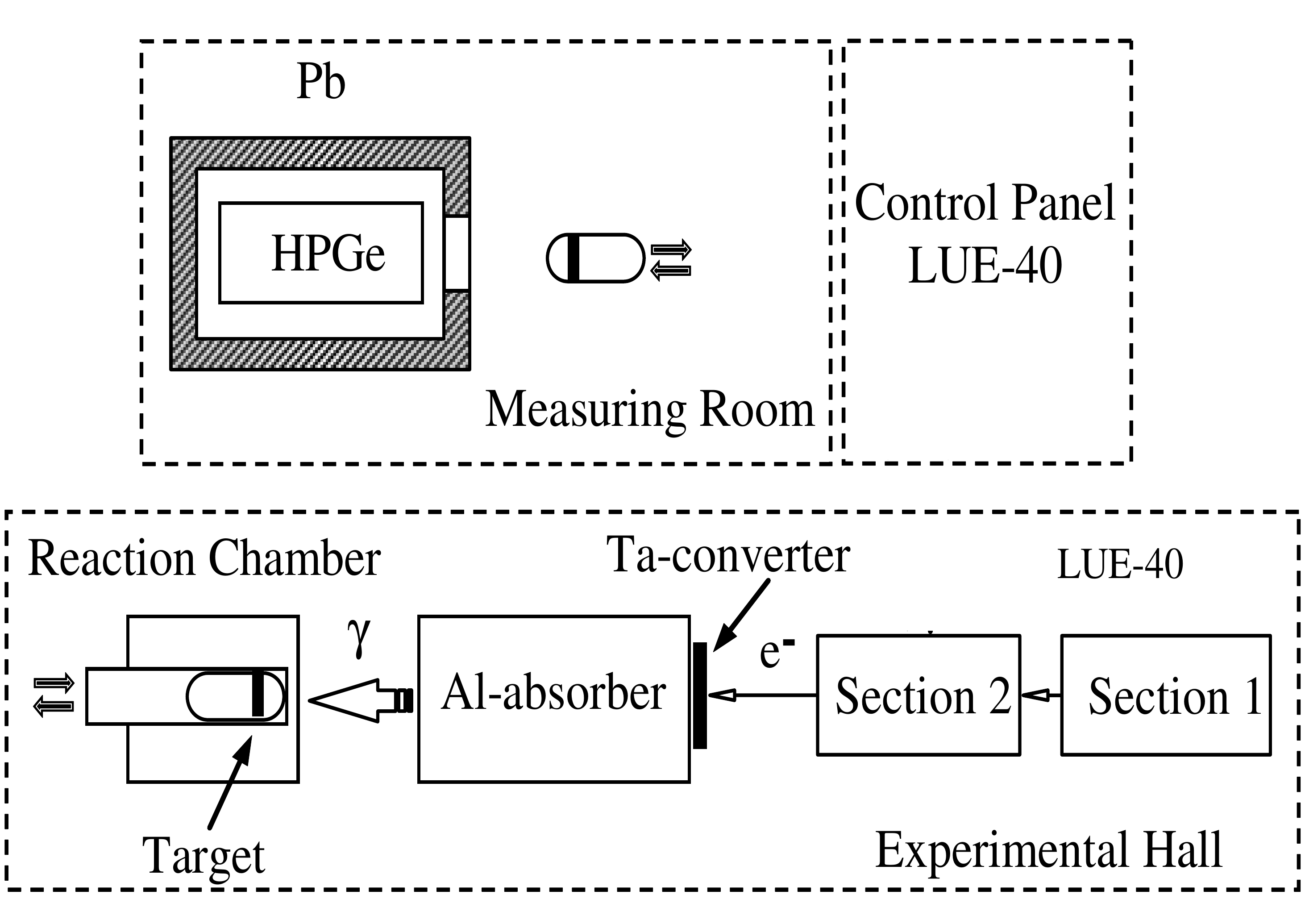}}
   	\caption{Schematic block diagram of the experiment. The upper part shows the measuring room and the room for accelerator performance control. The lower part shows (from right to left) two sections of the accelerator LUE-40; Ta converter; Al absorber; bombardment chamber.}
   	\label{fig1}
   \end{figure}

The block diagram of the experiment is shown in Fig.~\ref{fig1}, similar to \cite{21}. Electrons from the NSC KIPT electron linear accelerator LUE-40 \cite{23,24} with an initial energy $E_{\rm{e}}$ were incident on the target-convertor made of natural tantalum with cross dimensions of 20 mm x 20 mm and a thickness of 1.05 mm. To clean the flux of bremsstrahlung photons from electrons, a cylindrical Al absorber with a diameter of 100 mm and a length of 150 mm was used.

The bremsstrahlung $\gamma$-flux were calculated using the open certified code GEANT4.9.2 \cite{25} with due regard to the real geometry of the experiment, where spatial and energy distributions of the electron beam were taken into account. In addition, the bremsstrahlung $\gamma$-flux was controlled by the yield of the $^{100}\rm{Mo}(\gamma,n)^{99}\rm{Mo}$ reaction. For this purpose, simultaneously and placed close by the target under study, the natural molybdenum target-witness was exposed.
   
The targets made of natural tantalum and molybdenum with a diameter of 8~mm were placed in an aluminum capsule and, using a pneumatic transport system, were transported to the irradiation site and back to the detector. After the end of the irradiation, the capsule with the targets was delivered to the measurement room and within 150--200 sec the sample under study was removed from the capsule and placed to measure the induced $\gamma$-activity. This made it possible to experimentally obtain data on the yield of the reaction  $^{181}\rm{Ta}(\gamma,3n)^{178g}\rm{Ta}$, which has a relatively short half-life $T_{1/2} = 9.31\pm0.03$ min \cite{0}.   

\begin{figure*}[]
\begin{minipage}[h]{0.95\linewidth}
\center{\includegraphics[width=1\linewidth]{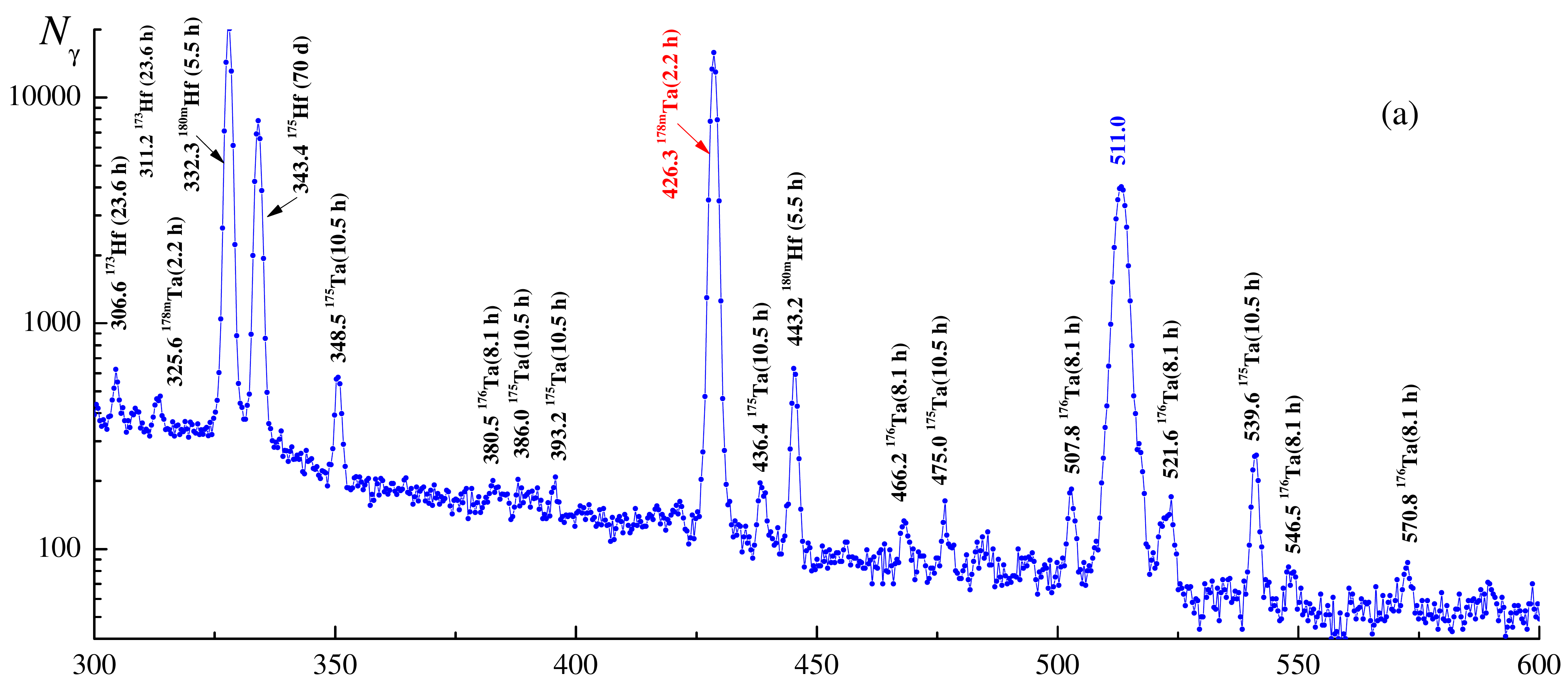}} \\
\end{minipage}
\vfill
\begin{minipage}[h]{0.95\linewidth}
\center{\includegraphics[width=1\linewidth]{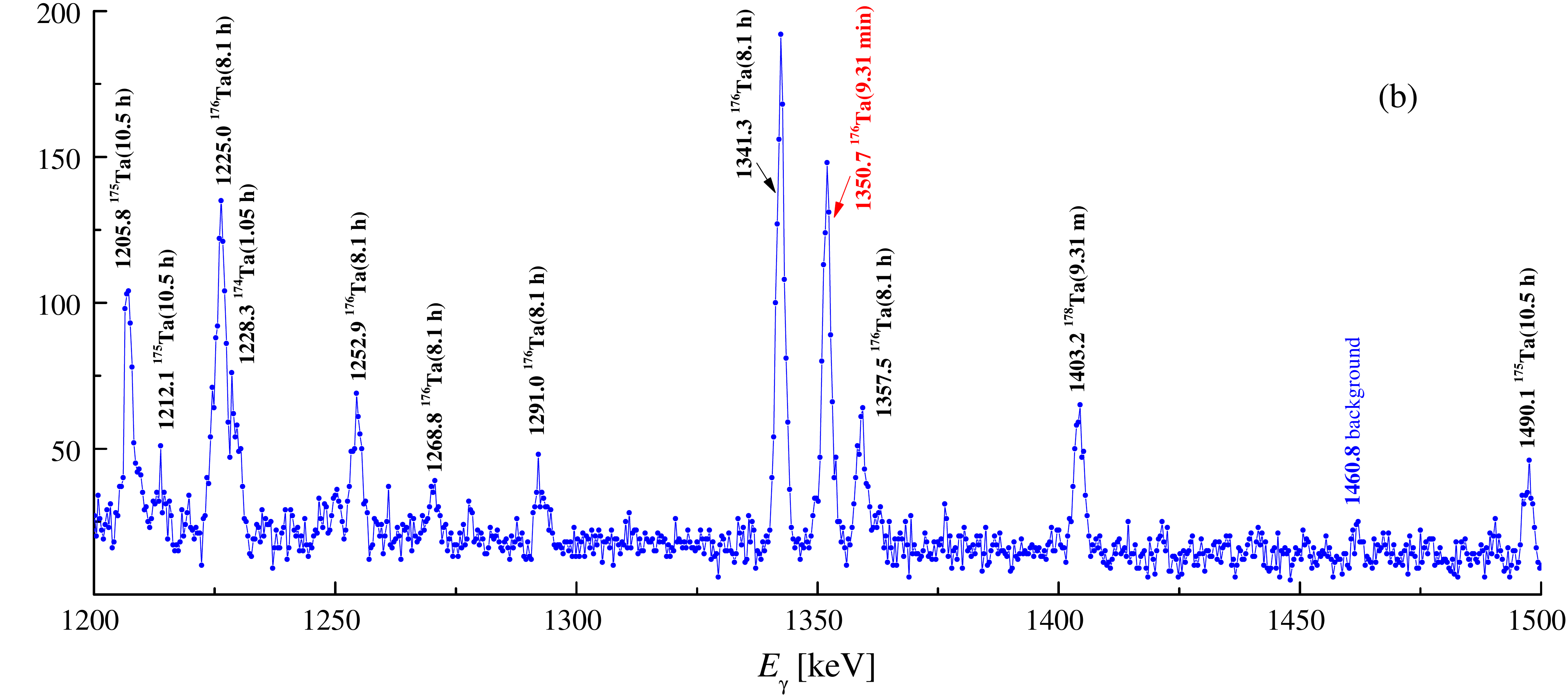}} \\
\end{minipage}
	\caption{Fragment of  $\gamma$-radiation spectrum in the energy range  $300 \leq E_{\gamma} \leq 600$~keV and  $1200 \leq E_{\gamma} \leq 1500$~keV from the tantalum target of mass 42.928~mg after exposure to the bremsstrahlung $\gamma$-flux at $E_{\rm{\gamma max}}$ = 80.7 MeV, $t_{\rm{irr}}$ = 30 min.}
	\label{fig2}
\end{figure*}

In the experiments, ten pairs of natural Ta/Mo samples were exposed to radiation at different end-point bremsstrahlung energies  $E_{\rm{\gamma max}}$ in the range from 35 to 80 MeV. The masses of Ta target and Mo target were, respectively, $\thicksim$43~mg and $\thicksim$60~mg. The time of irradiation $t_{\rm{irr}}$ and the time of residual activity spectrum measurement $t_{\rm{meas}}$ were both 30~min. To exemplify, Fig.~\ref{fig2} shows two fragments of  $\gamma$-radiation spectrum from the tantalum target in the energy ranges $300 \leq E_{\gamma} \leq 600$~keV and $1200 \leq E_{\gamma} \leq 1500$ keV in which the used in the work  $\gamma$-lines of the $^{178}$Ta nucleus were located.

For $\gamma$-radiation registration, a semiconductor HPGe detector (Canberra GC-2018) was used with resolutions of 1.8~keV and 0.8~keV (FWHM) for the energies $E_{\gamma}$ = 1332 keV and 122 keV, respectively. The absolute registration efficiency of the detector was calibrated with a standard set of $\gamma$-ray sources $^{22}$Na, $^{60}$Co, $^{133}$Ba, $^{137}$Cs, $^{152}$Eu and $^{241}\!\rm{Am}$.

  \begin{table*}[]
      	\caption{\label{tab1} Spectroscopic data \cite{0} on the nuclei-products from the reactions $^{181}\rm{Ta}(\gamma,3n)^{178\rm{m,g}}\rm{Ta}$ and the monitoring reaction $^{100}\rm{Mo}(\gamma,n)^{99}\rm{Mo}$}
      	\centering
      	\begin{tabular}{cccccc}
      		\hline	\vspace{1ex}
      		 \begin{tabular}{c} Nuclear \\  reaction \end{tabular} & $E_{\rm{th}}$,~MeV & \begin{tabular}{c} $J^\pi$ of  \\ nucleus-product
\end{tabular} & $T_{1/2}$ & $E_{\gamma}$,~keV & $I_{\gamma}$, \% \\ 	\hline
&&&&&\\ 	   
$^{181}\rm{Ta}(\gamma,3n)^{178g}\rm{Ta}$ & 22.21 & $1^+$ & 9.31 (3) min & 1350.68 (3) & 1.18 (3) \\
$^{181}\rm{Ta}(\gamma,3n)^{178m}\rm{Ta}$ & 22.24 & $(7)^-$ & 2.36 (8) h & 426.383 (6) & 97.0 (13) \\ \hline
$^{100}\rm{Mo}(\gamma,n)^{99}\rm{Mo}$   & 8.29   & $1/2^+$  & 65.94 (1) h & 739.50 (2) & 12.13 (12)  \\ 
	\hline
      	\end{tabular}	        
  \end{table*} 

  To investigate the reactions of interest, the residual activity $\gamma$-spectrum of the irradiated target was analyzed, and $\triangle A$  -- the number of counts of $\gamma$-quanta in the full absorption peak were determined for the $\gamma$-lines corresponding to the nuclei-products $^{178\rm{g}}$Ta and $^{178\rm{m}}$Ta. Relying on the data from ref.~\cite{0}, Table~1 gives the parameters of both the reactions under study and the monitoring reaction, viz.,  the energy $E_{\gamma}$ and intensity $I_{\gamma}$ of the $\gamma$-lines in use.

The bremsstrahlung $\gamma$-flux monitoring by the $^{100}\rm{Mo}(\gamma,n)^{99}\rm{Mo}$ reaction yield was performed by comparing the experimentally obtained average cross-section values with the computation data. To determine the experimental $\langle{\sigma(E_{\rm{\gamma max}})}\rangle_{\rm{exp}}$ values it has used the yield for the $\gamma$-line of energy $E_{\gamma}$ = 739.50~keV and intensity $I_{\gamma}$ = 12.13\% (see Table~1). The average cross-section $\langle{\sigma(E_{\rm{\gamma max}})}\rangle_{\rm{th}}$ values were computed with the cross-sections $\sigma(E)$ from the TALYS1.95 code with default options. Details of the monitoring procedure can be found in \cite{21,26}.

The Ta converter and Al absorber, used in the experiment, generate neutrons that can cause the reaction $^{100}\rm{Mo}(n,2n)^{99}\rm{Mo}$. Calculations were made of the energy neutrons spectrum and the fraction of neutrons with energies above the threshold of this reaction, similarly \cite{27,28}. The contribution of the $^{100}\rm{Mo}(n,2n)^{99}\rm{Mo}$ reaction to the value of the induced activity of the $^{99}\rm{Mo}$  nucleus has been estimated and it has been shown that this contribution is negligible compared to the contribution of $^{100}\rm{Mo}(\gamma,n)^{99}\rm{Mo}$. 

When calculating the values of the average reaction cross-sections, it was assumed that all radioactive isotopes were formed only as a result of photonuclear reactions at $^{181}$Ta, since the content of $^{180\rm{m}}$Ta isomer in a natural mixture of tantalum is negligible (0.012\%). The self-absorption of $\gamma$-radiation from the reaction products in the target was calculated using the GEANT4.9.2 code and was taken into account in the calculations.

The uncertainty of measurements of the average cross-sections $\langle{\sigma(E_{\rm{\gamma max}})}\rangle$, $\langle{\sigma(E_{\rm{\gamma max}})}\rangle_{\rm{m}}$ and $\langle{\sigma(E_{\rm{\gamma max}})}\rangle_{\rm{g}}$ was determined as a squared sum of statistical and systematic errors. The statistical error in the observed $\gamma$-activity is mainly associated with the statistics calculation for the total absorption peak of the corresponding $\gamma$-line and is estimated in the range of 2--10\% for $E_{\gamma}$  = 1350.68 keV and up to 2\% for $E_{\gamma}$  = 426.383 keV. The systematic errors stem from the uncertainties in 1) irradiation time -- 0.25--0.5\%;  2) electron current -- 0.5\%; 3) $\gamma$-radiation detection efficiency $\sim$2-2.5\%, mainly due to the measuring error of $\gamma$-radiation sources; 4) half-life period $T_{1/2}$ of reaction products and the intensity of the analyzed $\gamma$-quanta $I_{\gamma}$ (see Table 1); 5) normalization of the experimental data to the yield of the monitoring reaction $^{100}\rm{Mo}(\gamma,n)^{99}\rm{Mo}$ -- 2.5\%; 6) the GEANT4.9.2 computational error for the bremsstrahlung $\gamma$-flux $\sim$1.5\%.

It should be noted that the systematic error in the  $^{100}\rm{Mo}(\gamma,n)^{99}\rm{Mo}$ reaction yield is associated with 3 fundamentally unremovable errors, each $\sim$1\%: unidentified isotopic composition of natural molybdenum, uncertainty in the intensity of the used $\gamma$-line $I_{\gamma}$ \cite{0} and the statistical error in determining the area under the peak of the normalizing $\gamma$-line. The calculations used the value of the isotopic abundance of the $^{100}\rm{Mo}$ nucleus from \cite{25}, which is 9.63\%.

Thus, the experimental error of the obtained average cross-sections is within 6--11\%. 
   
      \section{CROSS-SECTIONS FOR THE REACTION $^{181}\rm{Ta}(\gamma,3n)^{178\rm{m,g}}\rm{Ta}$ FROM CODE TALYS1.95}
      \label{Cross-section}   
  
The theoretical values of the total $\sigma(E)$ and partial cross-sections of the $^{181}\rm{Ta}(\gamma,3n)^{178}\rm{Ta}$ reaction for monochromatic photons are taken from the TALYS1.95 code \cite{20}, which is installed on Linux Ubuntu-20.04. The calculations were performed for different level density ($LD$) models. There are 3 phenomenological level density models and 3 options for microscopic level densities:

$LD 1$: Constant temperature + Fermi gas model. In this model introduced by Gilbert and Cameron \cite{29}, the excitation energy range is divided into a low energy part from $E_0$ up to a matching energy $E_{\rm{M}}$, where the so-called constant temperature law applies and a high energy part above, where the Fermi gas model applies. 

$LD 2$: Back-shifted Fermi gas model. In the Back-shifted Fermi gas Model \cite{30}, the pairing energy is treated as an adjustable parameter and the Fermi gas expression is used down to $E_0$.

$LD 3$: Generalized superfluid model (GSM). The model takes superconductive pairing correlations into account according to the Bardeen-Cooper-Schrieffer theory. The phenomenological version of the model \cite{31,32} is characterized by a phase transition from a superfluid behavior at low energy, where pairing correlations strongly influence the level density, to a high energy region which is described by the Fermi gas model. The GSM thus resembles the constant temperature model to the extent that it distinguishes between low energy and a high energy region, although for the GSM this distinction follows naturally from the theory and does not depend on specific discrete levels that determine matching energy. Instead, the model automatically provides a constant temperature-like behavior at low energies.

$LD 4$: Microscopic level densities (Skyrme force) from Goriely’s tables. Using this model allows reading tables of microscopic level densities from RIPL database. These tables were computed by S. Gorielyon based on Hartree-Fock calculations for excitation energies up to 150~MeV and for spin values up to $I$ = 30. 

$LD 5$: Microscopic level densities (Skyrme force) from Hilaire’s combinatorial tables. The combinatorial model includes a detailed microscopic calculation of the intrinsic state density and collective enhancement. The only phenomenological aspect of the model is a simple damping function for the transition from spherical to deformed.  

$LD 6$: Microscopic level densities (temperature-dependent HFB, Gogny force) from Hilaire’s combinatorial tables.

\begin{figure}[]
\begin{minipage}[h]{0.99\linewidth}
{\includegraphics[width=1\linewidth]{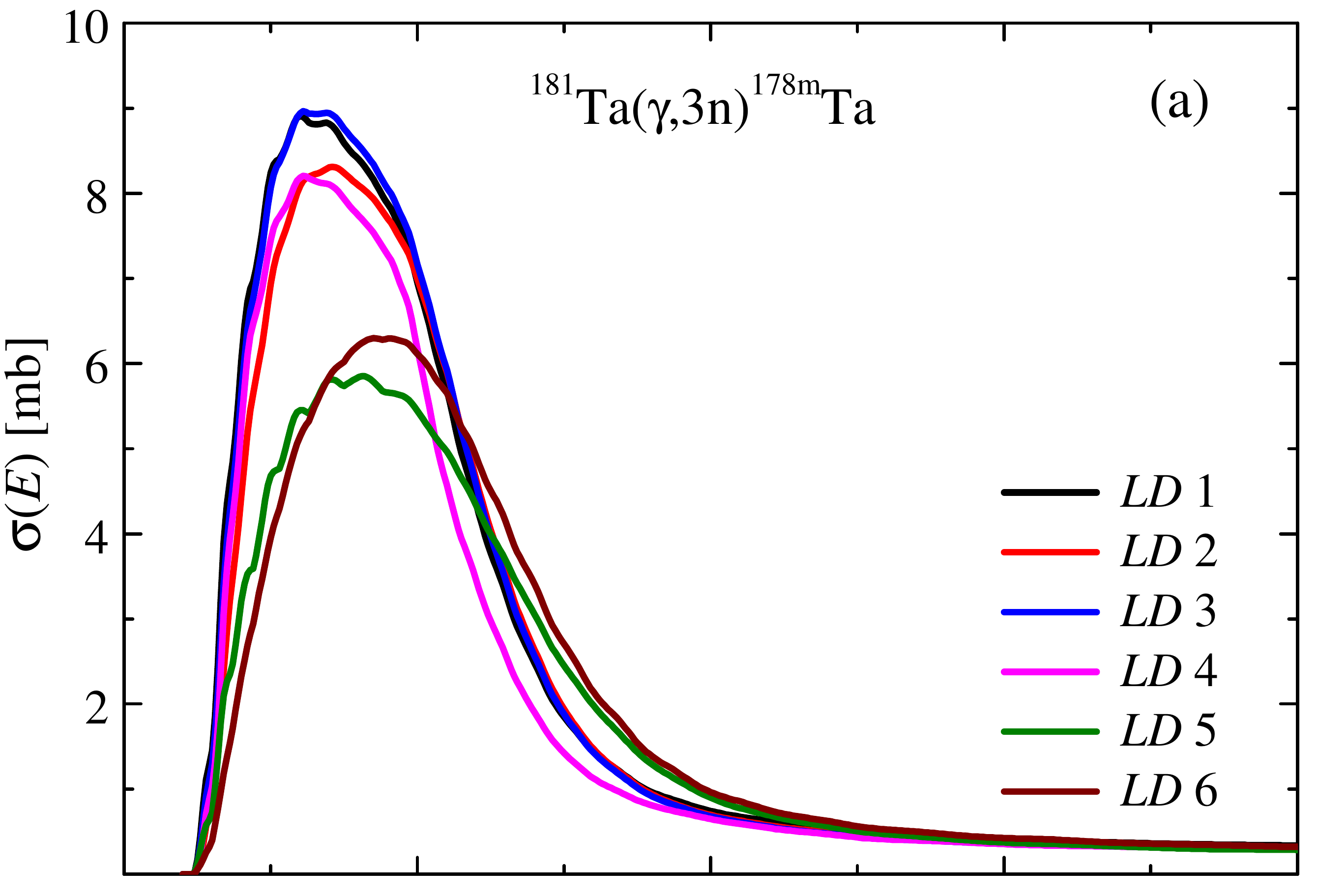}} \\
\end{minipage}
\vfill
\begin{minipage}[h]{0.99\linewidth}
{\includegraphics[width=1\linewidth]{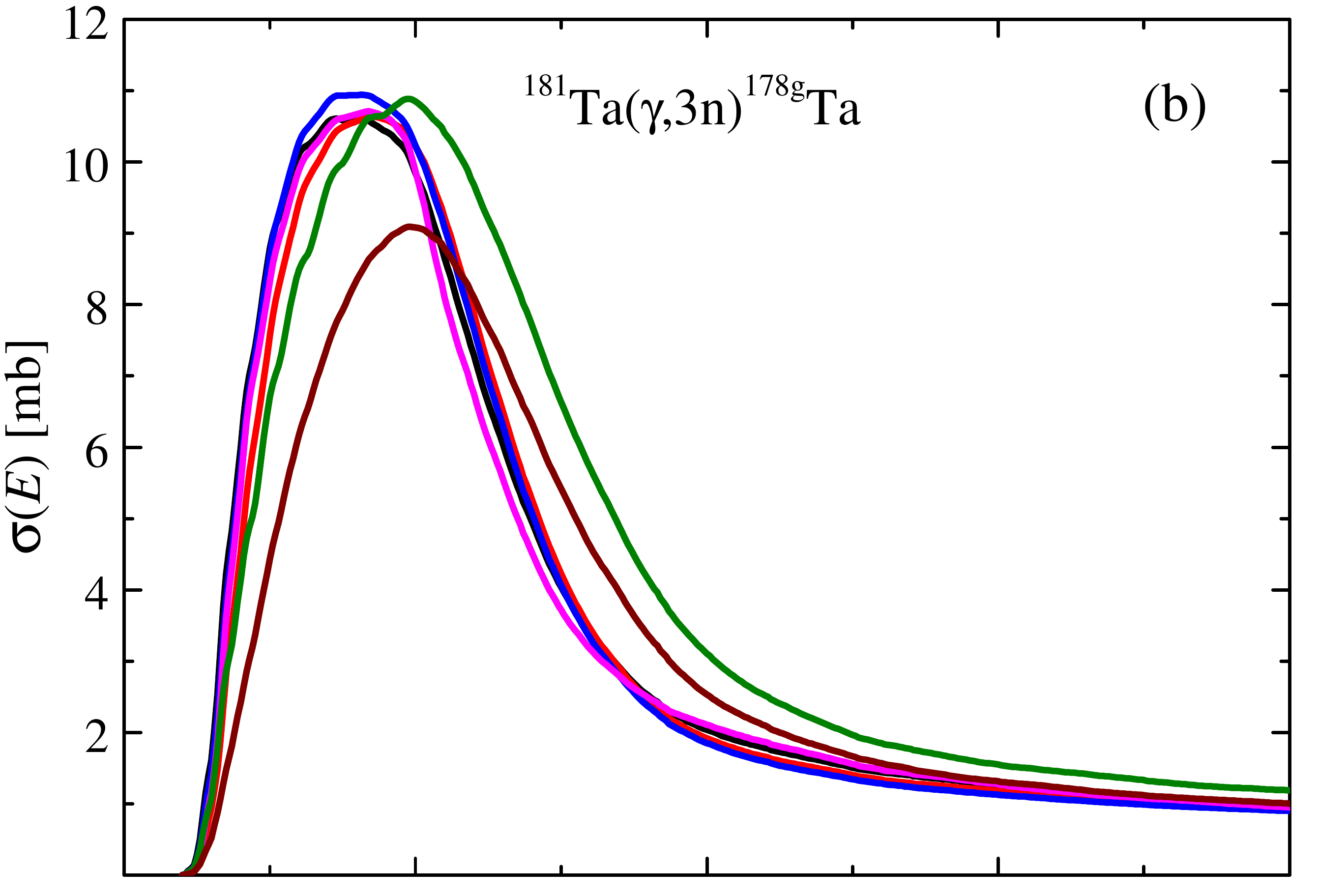}} \\
\end{minipage}
\vfill
\begin{minipage}[h]{0.99\linewidth}
{\includegraphics[width=1\linewidth]{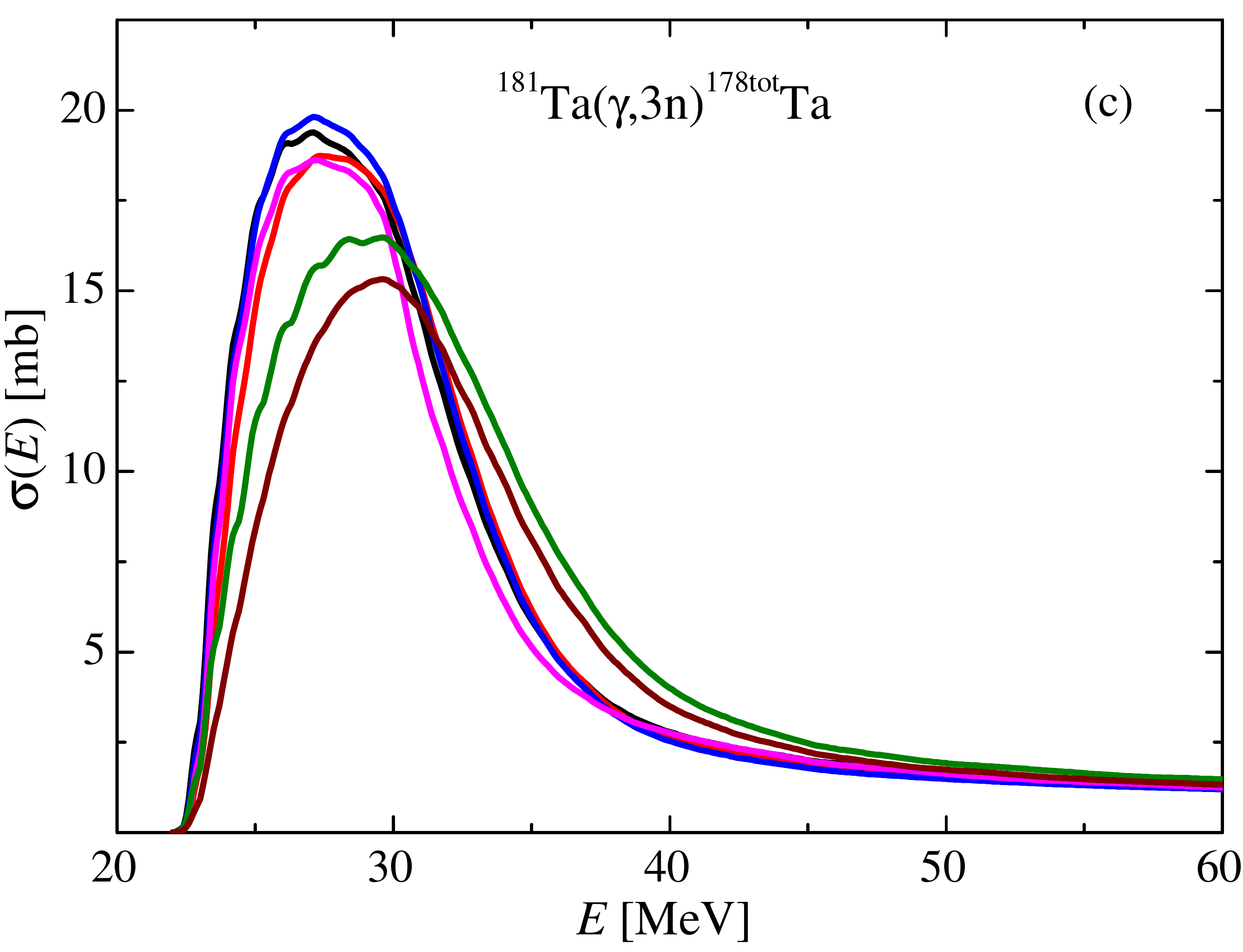}} \\
\end{minipage}
	\caption{Partial and total cross-sections $\sigma(E)$ for the reaction $^{181}\rm{Ta}(\gamma,3n)^{178}\rm{Ta}$ from the TALYS1.95 code for different models of level density  $LD$ 1-6. a) -- cross-section of nucleus formation in the isomeric state, b) -- in the ground state, c) -- total cross-section.}
	\label{fig3}
\end{figure}

Calculations of the total and partial cross-sections in the TALYS1.95 code for the $^{181}\rm{Ta}(\gamma,3n)^{178}\rm{Ta}$ reaction for different level density models are shown in Figs.~\ref{fig3} a, b, c. It can be seen from the figures that the behavior of the cross-sections on energy for the $LD$ 5 and $LD$ 6 models is somewhat different from the others: the positions of the cross-section maxima are shifted towards higher energy. Also, the height of the maxima of these curves in the case of calculations for the formation of a nucleus in the isomeric state is significantly lower (by 1.5 times) than for other $LD$ models. This leads to a decrease in the values of the total cross-section for the formation of the $^{178}$Ta nucleus calculated in the $LD$ 5 and $LD$ 6 models.

    \section{CALCULATION OF THE BREMSSTRAHLUNG FLUX-AVERAGE CROSS-SECTIONS AND THE ISOMERIC RATIO OF THE REACTION PRODUCTS}
      \label{Calc of ave}   

The values of the cross-sections  $\langle{\sigma(E_{\rm{\gamma max}})}\rangle$, averaged over the bremsstrahlung flux of $\gamma$-quanta $W(E,E_{\rm{\gamma max}})$ from the threshold $E_{\rm{th}}$ of the reaction under consideration to the boundary energy of the spectrum $E_{\rm{\gamma max}}$, were calculated using the cross-sections $\sigma(E)$ from the TALYS1.95 code [26]. The calculation of the bremsstrahlung flux-averaged cross-section  $\langle{\sigma(E_{\rm{\gamma max}})}\rangle$ was performed according to the formula: 
         
\begin{equation}\label{form1}
\langle{\sigma(E_{\rm{\gamma max}})}\rangle = \frac
{\int\limits_{E_{\rm{th}}}^{E_{\rm{\gamma max}}}\sigma(E)\cdot W(E,E_{\rm{\gamma max}})dE}
{\int\limits_{E_{\rm{th}}}^{E_{\rm{\gamma max}}}W(E,E_{\rm{\gamma max}})dE}.
\end{equation}

These calculated values of the average cross-sections were compared with the experimental values obtained in the work, which were calculated by the formula:
\begin{equation}\begin{split}
\langle{\sigma(E_{\rm{\gamma max}})}\rangle = \; \; \; \; \; \; \; \; \; \; \; \; \; \; \; \; \; \; \; \; \; \; \; \; \; \; \; \; \; \; \; \;\; \; \; \; \; \; \; \; \; \; \; \; \; \; \; \;\; \; \; \; \; \; \; \; \; \; \; \; \; \; \; \; \; \; \; \; \; \;   \\
\frac{\lambda \triangle A}{N_x I_{\gamma} \ \varepsilon \Phi(E_{\rm{\gamma max}}) (1-\exp(-\lambda t_{\rm{irr}}))\exp(-\lambda t_{\rm{cool}})(1-\exp(-\lambda t_{\rm{meas}}))},
\label{form2}
\end{split}\end{equation}
 $\lambda$ is the decay constant \mbox{($\rm{ln}2/\textit{T}_{1/2}$)},  where $\triangle A$ is the number of counts of $\gamma$-quanta in the full absorption peak (for the $\gamma$-line of the investigated reaction), $N_x$ is the number of target atoms, $I_{\gamma}$ is the  intensity of the analyzed $\gamma$-quanta, $\varepsilon$ is the absolute detection efficiency for the analyzed $\gamma$-quanta energy, $ {\rm{\Phi}}(E_{\rm{\gamma max}}) = {\int\limits_{E_{\rm{th}}}^{E_{\rm{\gamma max}}}W(E,E_{\rm{\gamma max}})dE}$  is the sum of bremsstrahlung quanta in the energy range from the reaction threshold $E_{\rm{th}}$ up to $E_{\rm{\gamma max}}$, $t_{\rm{irr}}$, $t_{\rm{cool}}$ and $t_{\rm{meas}}$ are the irradiation time, cooling time and measurement time, respectively. In more detail, all the calculation procedures required to determine $\langle{\sigma(E_{\rm{\gamma max}})}\rangle$ are described in \cite{19,21,26}. 

If the product nucleus has an isomeric state, the value of the total averaged cross-section $\langle{\sigma(E_{\rm{\gamma max}})}\rangle_{\rm{tot}}$ (hereinafter $\langle{\sigma(E_{\rm{\gamma max}})}\rangle$) of the reaction under study is calculated as the sum of $\langle{\sigma(E_{\rm{\gamma max}})}\rangle_{\rm{m}}$ and $\langle{\sigma(E_{\rm{\gamma max}})}\rangle_{\rm{g}}$, respectively, of the average cross-sections for the population of the isomeric  and ground states, each with its reaction threshold $E_{\rm{th}}$ (see Table~1).

The isomeric ratio of the reaction products is defined as the ratio of the cross-section  $\sigma_{\rm{m}}(E)$ of the formation of the nucleus in the metastable (usually high-spin) state to the cross-section $\sigma_{\rm{g}}(E)$ of the nucleus in the ground state. This definition makes it possible to estimate the degree of the population of the metastable state concerning the ground state of the product nucleus.

In the case of experiments on the bremsstrahlung beam of $\gamma$-quanta, the isomeric ratio is defined as the ratio of yields or as the ratio of the average cross-sections for the formation of reaction products in the metastable and ground states. The expression for $d(E_{\rm{\gamma max}})$ in terms of average cross-sections can be written as:

\begin{equation}\label{form3}
d(E_{\rm{\gamma max}}) = 
{\langle{\sigma(E_{\rm{\gamma max}})}\rangle_{\rm{m}}} / 
{\langle{\sigma(E_{\rm{\gamma max}})}\rangle_{\rm{g}}}.
\end{equation}

 \section{RESULTS AND DISCUSSION}
 \label{RES AND DISC} 
\subsection{The bremsstrahlung flux-averaged cross-sections  $\langle{\sigma(E_{\rm{\gamma max}})}\rangle$ for the reaction $^{181}\rm{Ta}(\gamma,3n)^{178}\rm{Ta}$}
 \label{subsec1}     

The obtained experimental data on the bremsstrahlung flux-averaged cross-sections  $\langle{\sigma(E_{\rm{\gamma max}})}\rangle$, $\langle{\sigma(E_{\rm{\gamma max}})}\rangle_{\rm{g}}$ and $\langle{\sigma(E_{\rm{\gamma max}})}\rangle_{\rm{m}}$ for the $^{181}\rm{Ta}(\gamma,3n)^{178}\rm{Ta}$ reactions in the end-point energy range of bremsstrahlung $\gamma$-quanta $E_{\rm{\gamma max}}$ = 35--80 MeV are presented in Figs. 4 a, b, c. These figures also show the earlier results of our study \cite{19} for the considered reaction (at $E_{\rm{\gamma max}}$ = 80--95 MeV). Within the limits of experimental errors, the data of this work and from \cite{19} are in good agreement.

\begin{figure}[]
\begin{minipage}[h]{0.99\linewidth}
{\includegraphics[width=1\linewidth]{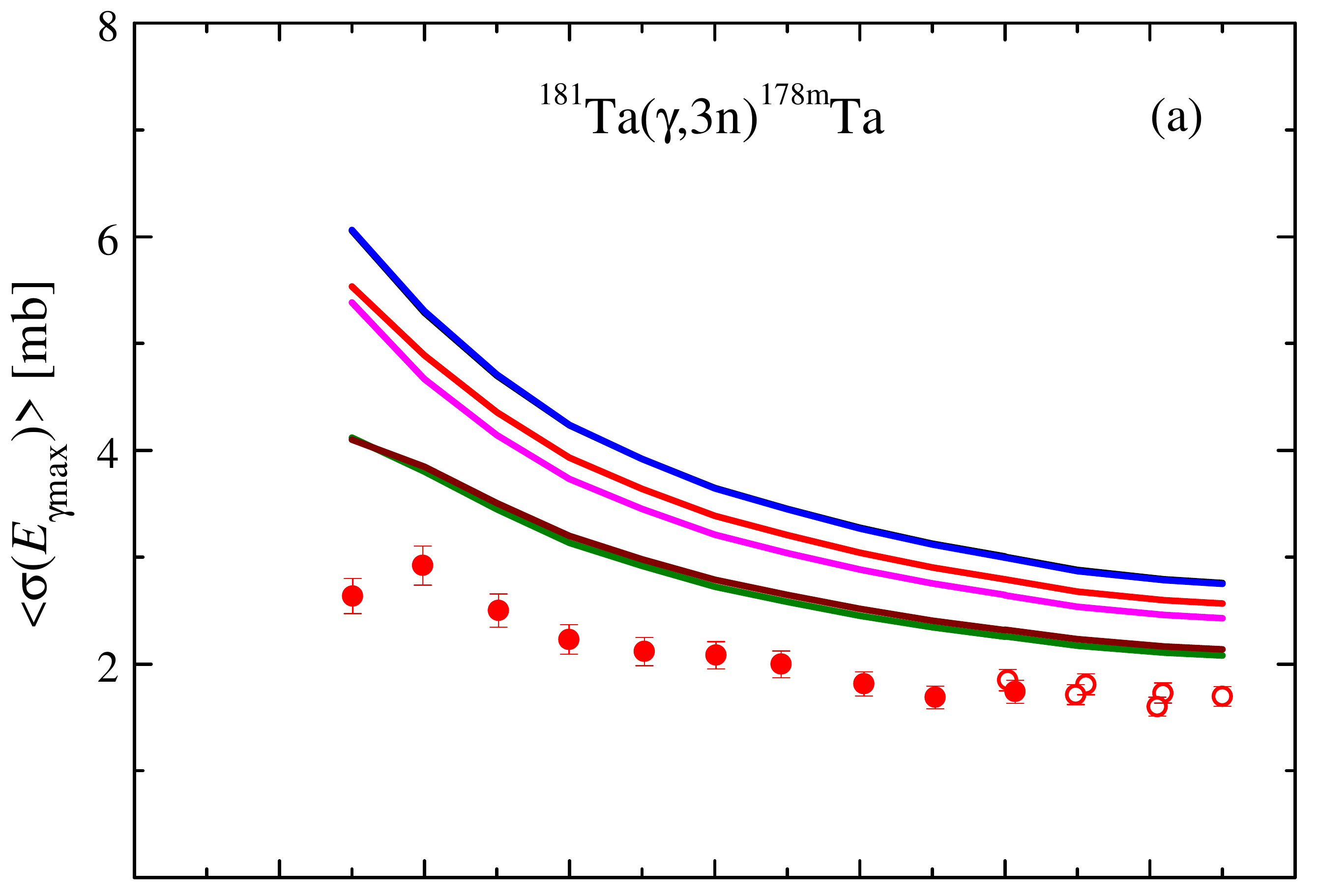}} \\
\end{minipage}
\vfill
\begin{minipage}[h]{0.99\linewidth}
{\includegraphics[width=1\linewidth]{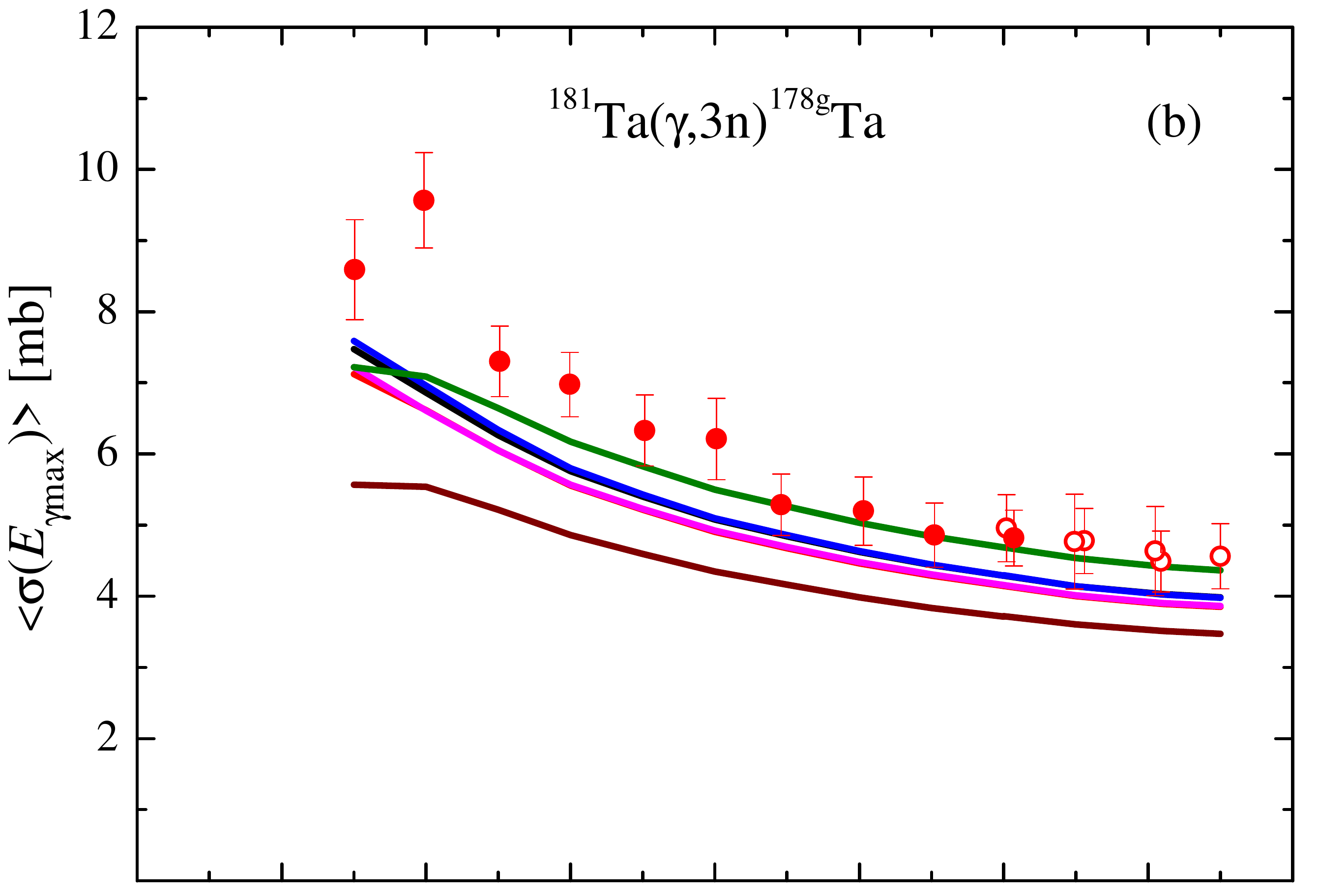}} \\
\end{minipage}
\vfill
\begin{minipage}[h]{0.99\linewidth}
{\includegraphics[width=1\linewidth]{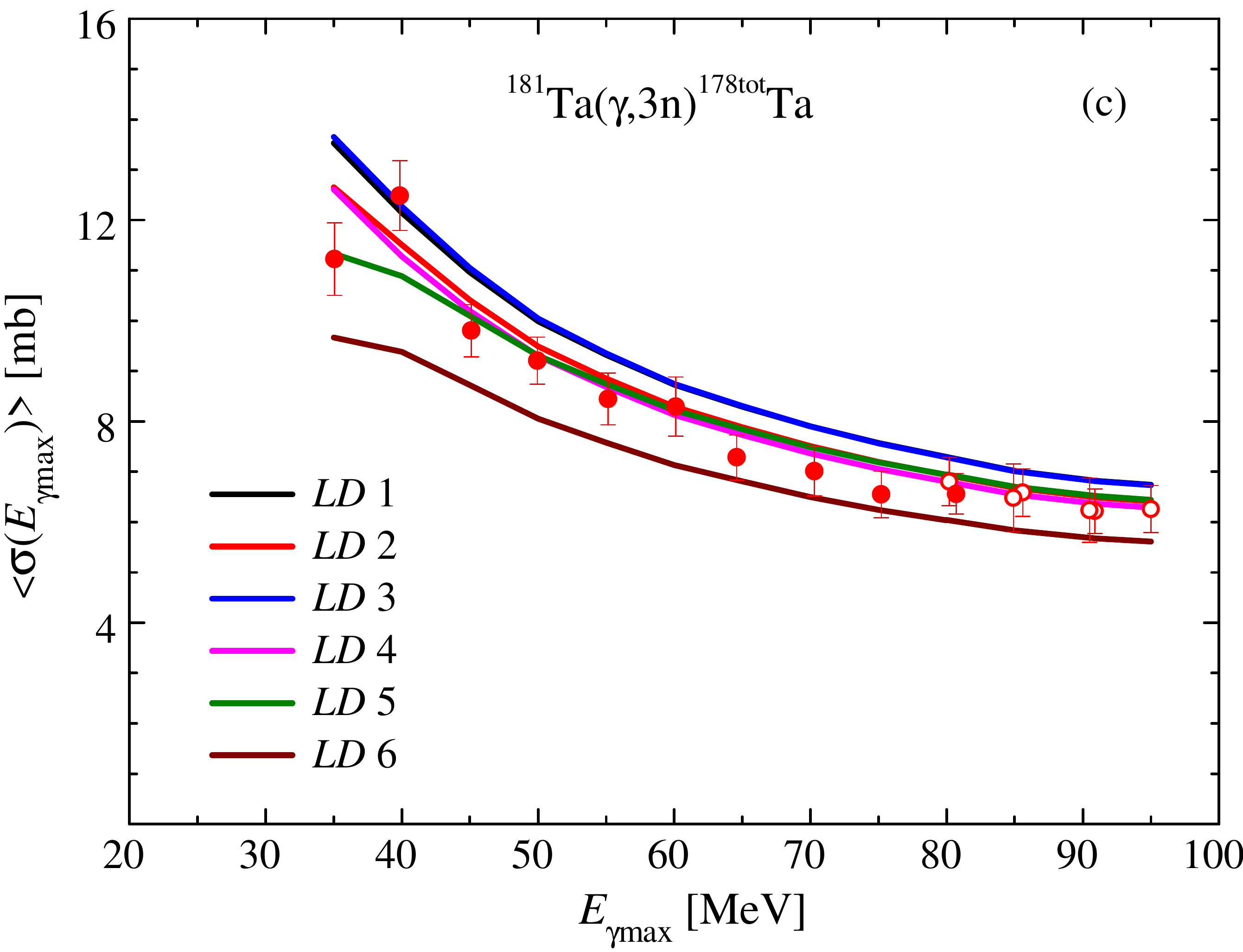}} \\
\end{minipage}
	\caption{The bremsstrahlung flux-averaged cross-sections partial  and total for the reaction $^{181}\rm{Ta}(\gamma,3n)^{178}\rm{Ta}$: a) -- the cross-section for the formation of  nucleus in the isomeric state, b) -- in the ground state, c) -- the total cross-section of the reaction. Points -- experimental results of this work (filled) and \cite{19} (empty), curves -- calculation using the TALYS1.95 code for different models of the $LD$ 1-6 level density.}
	\label{fig4}
\end{figure}

The theoretical values of the average cross-sections are obtained from eq.~(\ref{form1}) using bremsstrahlung fluxes corresponding to real experimental conditions and are shown in Fig.~\ref{fig4}. It can be seen from Fig.~\ref{fig4}a that the experimental cross-sections for the formation of a nucleus in the isomeric state $\langle{\sigma(E_{\rm{\gamma max}})}\rangle_{\rm{m}}$  are located below all theoretical curves, but closest to the calculations by models $LD$ 5 and 6. In the case of the cross-section $\langle{\sigma(E_{\rm{\gamma max}})}\rangle_{\rm{g}}$  the experimental values are higher than all calculations in the 35--60 MeV range, but at 65--95 MeV there is in good agreement with the calculation for the $LD$ 5 model.

As can be seen from Fig.~\ref{fig4}c, the theoretical values of the total averaged cross-section according to the $LD$ 1,3 and $LD$ 6 models differ by 20--30\%, forming a corridor in which all the experimental data $\langle{\sigma(E_{\rm{\gamma max}})}\rangle$ and calculations for $LD$ 2,4,5 are located. The insignificant difference between the theoretical values $\langle{\sigma(E_{\rm{\gamma max}})}\rangle$ for $LD$ 2, 4, and 5 (up to 5\% in the energy range of 45--95 MeV) does not allow one of them to be distinguished using experimental results.

In general, the analysis of the experimental values of $\langle{\sigma(E_{\rm{\gamma max}})}\rangle$, $\langle{\sigma(E_{\rm{\gamma max}})}\rangle_{\rm{m}}$ and $\langle{\sigma(E_{\rm{\gamma max}})}\rangle_{\rm{g}}$ shows that the best agreement with theoretical calculations using TALYS1.95 code was achieved for the $LD$ 5 model: Microscopic level densities (Skyrme force) from Hilaire's combinatorial tables.

 \subsection{Isomeric ratio $d(E_{\rm{\gamma max}})$ of the reaction products of the $^{181}\rm{Ta}(\gamma,3n)^{178m,g}\rm{Ta}$}
  \label{subsec2}

\begin{figure}[h]
	\center{\includegraphics[scale=0.3]{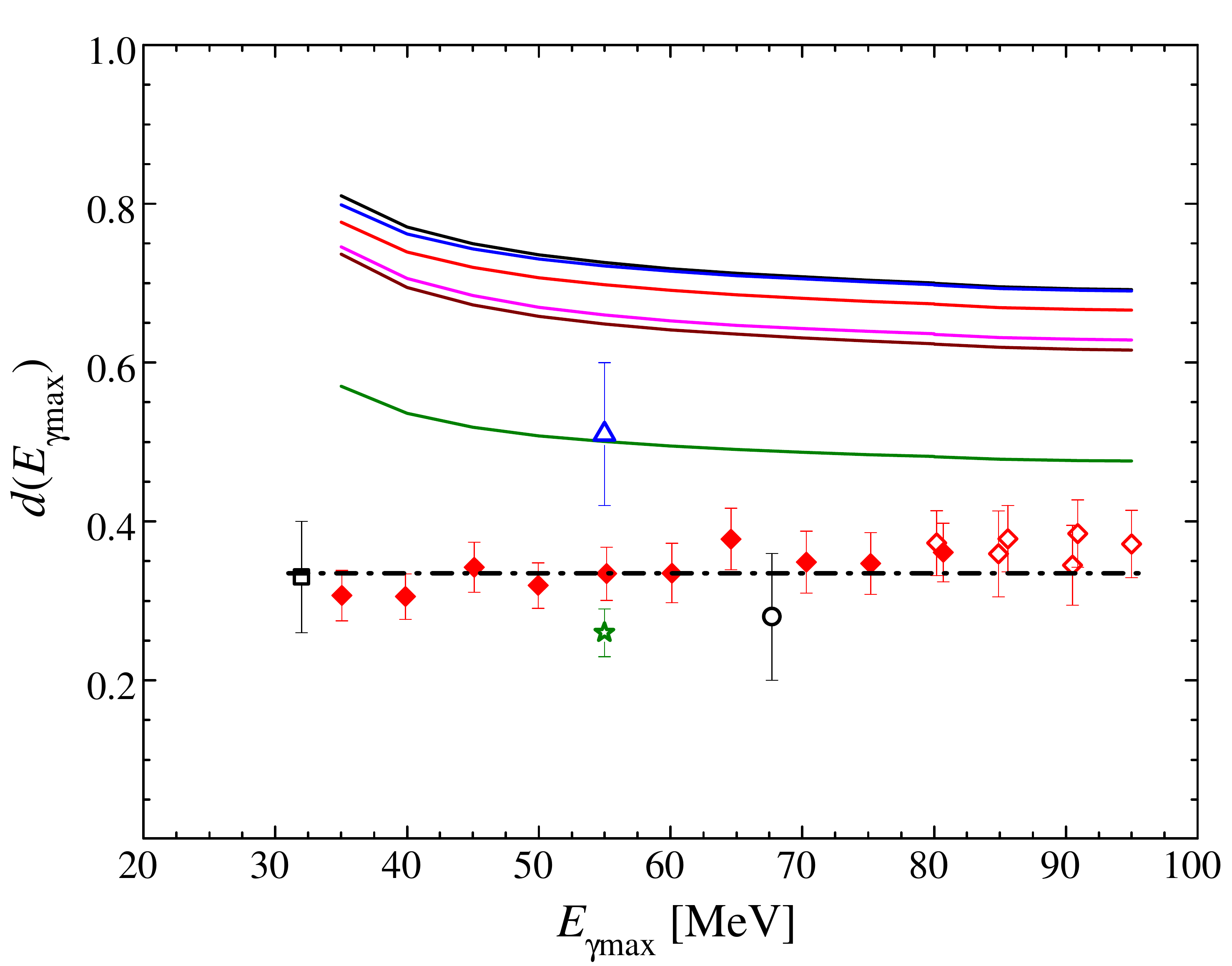}}
	\caption{Isomeric ratios $d(E_{\rm{\gamma max}})$ of the $^{181}\rm{Ta}(\gamma,3n)^{178m,g}\rm{Ta}$ reaction products. Experimental data: rhombuses -- the result of this work (filled) and from \cite{19} (empty), star -- \cite{14}, circle -- \cite{15}, square -- \cite{17}, triangle -- \cite{18}. Solid lines -- calculations of $d(E_{\rm{\gamma max}})$ using TALYS1.95 for the models $LD$ 1-6 (the lines are marked as in Fig.~4), dash-dotted line -- approximation of all experimental data by horizontal line (0.335$\pm$0.008).}
	\label{fig5}
\end{figure}

In Fig.~\ref{fig5} shows the experimental and theoretical values of the isomeric ratio $d(E_{\rm{\gamma max}})$ of the reaction products $^{181}\rm{Ta}(\gamma,3n)^{178m,g}\rm{Ta}$, calculated by eq.~(\ref{form3}) using the obtained $\langle{\sigma(E_{\rm{\gamma max}})}\rangle_{\rm{m}}$ and $\langle{\sigma(E_{\rm{\gamma max}})}\rangle_{\rm{g}}$.

The experimental values $d(E_{\rm{\gamma max}})$ found in this work coincides within the error with the results in the range $E_{\rm{\gamma max}}$ = 80--95 MeV \cite{19}. Analysis of the entire set of our experimental values shows that $d(E_{\rm{\gamma max}})$ are grouped around a constant value 0.343$\pm$0.007. This result refines the previously obtained value of $d(E_{\rm{\gamma max}})$ equal to 0.37$\pm$0.02 \cite{19}. Attention is drawn to the tendency for an insignificant decrease in the isomeric ratio with decreasing energy $E_{\rm{\gamma max}}$. 

The value of the isomeric ratio $d(E_{\rm{\gamma max}})$  makes it possible to estimate the degree of the population of the metastable state concerning the ground state of the reaction product nucleus. The found value of the isomeric ratio shows that the probability of the formation of a nucleus in the ground state of $^{178\rm{g}}$Ta ($J^{\pi} = 1^+$) exceeds the probability of the formation of the isomeric state of the $^{178\rm{m}}$Ta ($J^{\pi} = (7)^-$) by about 3 times.

The experimental values of the isomeric ratio $d(E_{\rm{\gamma max}})$ of the reaction products $^{181}\rm{Ta}(\gamma,3n)^{178m,g}\rm{Ta}$ available in the literature \cite{15,14,17,18} make it possible to expand the energy range to 24--95 MeV. In 
\cite{17}, the values of $d(E_{\rm{\gamma max}})$ were obtained for energies of 24--32 MeV, and the constancy of the isomer ratio  
equal to 0.33$\pm$0.07 was shown (in Fig.~\ref{fig5} they are shown only for 32 MeV). A joint analysis of the values from \cite{15,14,19,17} and this work shows that the experimental values of $d(E_{\rm{\gamma max}})$ are grouped near a constant value 
equal to 0.335$\pm$0.008. Since the value $d(E_{\rm{\gamma max}})$ from \cite{18} is significantly higher than all experimental results, it was not used in the analysis. 

In \cite{33}, the isomeric ratios of the reaction products $^{181}\rm{Ta}(p,p3n)^{178m,g}\rm{Ta}$ were obtained at proton energies of 100, 145, 200, 350, and 500 MeV using by off-line $\gamma$-spectroscopy. It is shown that the obtained isomeric ratios are constant and close to $d(E) \sim$ 0.37--0.38 at 100--500 MeV. This value is consistent with the results of this work and coincides with an energy of 95--100 MeV. We emphasize the experimental fact that, despite the different excitation mechanisms of the $^{181}$Ta nucleus, the isomeric ratio of the reaction products $^{178\rm{m,g}}$Ta is the same.

In Fig.~\ref{fig5} the theoretical values of the isomeric ratio  $d(E_{\rm{\gamma max}})$ calculated for the fluxes of bremsstrahlung $\gamma$-quanta for real experimental conditions are shown. All calculated curves for  $d(E_{\rm{\gamma max}})$ ($LD$ 1-6) have the same dependence on the energy $E_{\rm{\gamma max}}$: the isomeric ratio slightly increases with decreasing energy. The main difference between the calculated curves lies in the absolute value of the isomeric ratio. The calculation for the $LD$ 5 model lies below all other models.

The comparison shows that the default calculation ($LD$ 1) gives a value of $d(E_{\rm{\gamma max}})$ almost two times higher than the experiment, and the best agreement with the experiment was achieved for the case of the $LD$ 5 model.

Note the different behavior of the calculated and experimental energy dependence of isomeric ratios. The calculated value of $d(E_{\rm{\gamma max}})$ increases with decreasing energy $E_{\rm{\gamma max}}$ and the available experimental data tend to insignificantly (within the limits of experimental errors) decrease.

\section{Conclusions}
\label{Concl}

The present work has been concerned with the investigation of the photoneutron reaction $^{181}\rm{Ta}(\gamma,3n)^{178m,g}\rm{Ta}$ at the boundary energies of bremsstrahlung spectra $E_{\rm{\gamma max}}$ = 35--80 MeV using the residual $\gamma$-activity method. As a result, the flux-averaged cross-sections $\langle{\sigma(E_{\rm{\gamma max}})}\rangle$, $\langle{\sigma(E_{\rm{\gamma max}})}\rangle_{\rm{m}}$ and $\langle{\sigma(E_{\rm{\gamma max}})}\rangle_{\rm{g}}$ of photoneutron reactions $^{181}\rm{Ta}(\gamma,3n)^{178m,g}\rm{Ta}$.

The theoretical values $\langle{\sigma(E_{\rm{\gamma max}})}\rangle$, $\langle{\sigma(E_{\rm{\gamma max}})}\rangle_{\rm{m}}$ and $\langle{\sigma(E_{\rm{\gamma max}})}\rangle_{\rm{g}}$ are calculated using $\sigma(E)$ from the TALYS1.95 code for different models of the $LD$ 1-6 level density. The comparison of the experimental averaged cross-sections with calculations shows that the best agreement was achieved for the $LD$ 5 model: Microscopic level densities (Skyrme force) from Hilaire's combinatorial tables.

The experimental values of isomeric ratios $d(E_{\rm{\gamma max}})$ of the reaction products were obtained in the energy range $E_{\rm{\gamma max}}$ = 35--80 MeV. The found $d(E_{\rm{\gamma max}})$ indicate a significant suppression of the population of the isomeric state of the $^{178\rm{m}}$Ta nucleus concerning the ground state of $^{178\rm{g}}$Ta (by about a factor of 3). The obtained $d(E_{\rm{\gamma max}})$ values agree with the literature data. All available experimental $d(E_{\rm{\gamma max}})$ \cite{15,14,19,17,18} in the range $E_{\rm{\gamma max}}$ = 24--95 MeV are grouped near a constant value of 0.335$\pm$0.008.

It is shown that the default calculation ($LD$ 1) gives the value of $d(E_{\rm{\gamma max}})$ almost two times higher than the experiment. The closest to the experiment is the calculation for the $LD$ 5 model.

The different behavior of energy dependences of the calculated and experimental isomeric ratios is observed. The calculated value of $d(E_{\rm{\gamma max}})$ increases with decreasing energy $E_{\rm{\gamma max}}$ and the available experimental data tend to slightly (within the experimental errors) decrease with decreasing energy $E_{\rm{\gamma max}}$.

\section*{Acknowledgment}
The authors would like to thank the staff of the linear electron accelerator LUE-40 NSC KIPT, Kharkiv, Ukraine, for their cooperation in the realization of the experiment.

\section*{Declaration of competing interest}\addcontentsline{toc}{section}{Declaration of competing interest}
\label{Decl}
The authors declare that they have no known competing financial interests or personal relationships that could have appeared to influence the work reported in this paper.




\begin{thebibliography}{00}
\bibitem{0}
S.Y.F. Chu, L.P. Ekstrom, R.B. Firestone, The Lund/LBNL, Nuclear Data
Search, Version 2.0, February 1999, WWW Table of Radioactive Iso-
topes, available from http://nucleardata.nuclear.lu.se/toi/.

\bibitem{1a}
R. Volpel, Nucl. Phys. A, 182: 411 (1972), doi.org/10.1016/0375-9474(72)90287-4.

\bibitem{2a}
H. Bartsch, K. Huber, U. Kneissl, and H. Krieger, Nucl. Phys. A, 256: 243 (1976), doi.org/10.1016/0375-9474(76)90106-8.

\bibitem{3a}
 D. Kolev, E. Dobreva, N. Nenov, and V. Todorov, Nucl. Instrum. Meth. Phys. Research A 356: 390 (1995), doi.org/10.1016/0168-9002(94)01319-5.

\bibitem{4a}
I.B. Haller, G. Rudstam, J. Inorg. Nucl. Chem. 19: 1 (1961), doi.org/10.1016/0022-1902(61)80038-9.

\bibitem{5a}
 D. Kolev, Appl. Radiat. Isot. 49: 989 (1998), doi.org/10.1016/S0969-8043(97)10094-X. 

\bibitem{1}
J.R. Huizenga and R. Vandenbosch, Phys. Rev. 120: 1305 (1960), doi.org/10.1103/PhysRev.120.1305. 

\bibitem{2}
 R. Vandenbosch and J.R. Huizenga, Phys. Rev. 120: 1313 (1960), doi.org/10.1103/PhysRev.120.1313. 

\bibitem{3}
H. A. Bethe, Rev. Mod. Phys. 9: 84 (1937), doi.org/10.1103/RevModPhys.9.69.

\bibitem{4}
 C. Bloch,  Phys. Rev. 93: 1094 (1954), doi.org/10.1103/PhysRev.93.1094. 

\bibitem{5}
K.J. Le Couteur and D.W. Lang, Nucl. Phys. 13: 32 (1959), doi.org/10.1016/0029-5582(59)90136-1. 

\bibitem{15}
B.S. Ishkhanov, V.N. Orlin, and S.Yu. Troschiev, Phys. Atom. Nucl. 75: 253 (2012) doi.org/10.1134/S1063778812020093.

\bibitem{5b}
 Experimental Nuclear Reaction Data (EXFOR) // https://www-nds.iaea.org/exfor/

\bibitem{6b}
 Data Center of Photonuclear Experiments, http://cdfe.sinp.msu.ru/.

\bibitem{6}
G.M.~Gurevich, L.E.~Lazareva, V.M.~Mazur, et al.,  Nucl. Phys. A  351: 257 (1981), doi.org/10.1016/0375-9474(81)90443-7.

\bibitem{7}
R.L.~Bramblett, J.T.~Caldwell, G.F.~Auchampaugh, and S.C.~Fultz, Phys. Rev. 129: 2723 (1963), doi.org/10.1103/PhysRev.129.2723.

\bibitem{8}
R.~Bergere, H.~Beil, A.~Veyssiere,  Nucl. Phys. A 121: 463 (1968), doi.org/10.1016/0375-9474(68)90433-8.

\bibitem{9}
	E.G.~Fuller, and M.S.~Weiss,  Phys. Rev. 112:  560 (1958), doi.org/10.1103/PhysRev.112.560.

\bibitem{10}
	O.V.~Bogdankevich, B.I.~Goryachev, V.A.~Zapevalov. Sov. Phys. JETP 42: 1502 (1962).

\bibitem{11}
 G.P.~Antropov, I.E.~Mitrofanov, B.S.~Russkikh. Bull. Acad. Sci. USSR. Phys. 31: 336 (1967).

\bibitem{12}
	B.S.~Ishkhanov, I.M.~Kapitonov, E.V.~Lazutin, et al., JETP Letters 54: 80 (1969).
\bibitem{13}
	S.N.~Belyaev, and V.P.~Sinichkin. in Proceedings of the 8th International Meeting on Beam dynamics and optimization. Saratov Univ. Publ., Saratov, 2002, 81.

\bibitem{16}
T. Kawano, Y.S. Cho, P. Dimitriou, et al.  Nucl. Data Sheets 163: 109 (2020), https://doi.org/10.1016/j.nds.2019.12.002.

\bibitem{14}
 V. Zheltonozhsky, M. Zheltonozhskaya, A.~Savrasov, and A.~Chernyaev, Book of Abstracts "LXX Int. Conf. ”NUCLEUS – 2020” online part, 12-17 October 2020, Saint Petersburg,  64.

\bibitem{19}
A.N. Vodin, O.S. Deiev, I.S. Timchenko, et al.,  EPJ A 57: 208 (2021), doi.org/10.1140/epja/s10050-021-00484-x, arXiv:2103.09859.

\bibitem{20}
A. Koning, and D. Rochman, Nucl. Data Sheets 113: 2841 (2012), TALYS -- based evaluated nuclear data library. https://tendl.web.psi.ch/tendl 2019/tendl2019.html.

\bibitem{17}
J.H. Carver and W. Turchinetz, The $(\gamma,2\rm{n})$ and  $(\gamma,3\rm{n})$ reactions in $^{181}$Ta // Research School of Physical Sciences, Australian National University, Canberra, September 1957, p. 613.

\bibitem{18}
H. Bartsch, K. Huber, U. Kneissl and H. Krieger, Nucl. Phys. A 256: 243 (1976), doi.org/10.1016/0375-9474(76)90106-8.

\bibitem{21}
A.N. Vodin, O.S. Deiev, I.S. Timchenko, et al., Probl. Atom. Sci. Tech. 3: 148 (2020). 

\bibitem{22}
H. Naik, G.N. Kim, R. Schwengner, et al.,  Nucl. Phys. A 916: 168 (2013), dx.doi.org/10.1016/ j.nuclphysa.2013.08.003.

\bibitem{23} 
A.N. Dovbnya, M.I. Aizatsky, V.N. Boriskin, et al., Probl. Atom. Sci. Tech. 2: 11 (2006).

\bibitem{24} 
 M.I. Aizatskyi, V.I. Beloglasov, V.N. Boriskin, et al., Probl. Atom. Sci. Tech. 3:60 (2014).

\bibitem{25} 
S. Agostinelli, J.R. Allison, K. Amako, et al., GEANT4-a simulation toolkit, Nucl. Instrum. Meth. Phys. Research A 506: 250 (2003), doi:10.1016/S0168-9002(03)01368-8,
http://GEANT4.9.2.web.cern.ch/GEANT4.9.2/. 

\bibitem{26}
A.N. Vodin, O.S. Deiev, V.Yu. Korda, et al., Nucl. Phys. A 1014:  122248 (2021), doi.org/10.1016/j.nuclphysa.2021.122248, arXiv:2101.08614.

\bibitem{27}
O.S. Deiev, G. L. Bochek, V. N. Dubina, et al., Probl. Atom. Sci. Tech. 121: 65 (2019).

\bibitem{28}
O.M. Vodin, O.S. Deiev, and S.M. Olejnik.   Probl. Atom. Sci. Tech. 124: 122 (2019).

\bibitem{29}
 A. Gilbert, and A.G.W. Cameron, Canad. Journ. Phys. 43: 1446 (1965).

\bibitem{30}
W. Dilg, W. Schantl, H. Vonach, and M. Uhl.  Nucl. Phys. A 217: 269 (1973), doi.org/10.1016/0375-9474(73)90196-6.

\bibitem{31}
V.G. Nedorezov, and Yu.N. Ranyuk. Photofission of nuclei behind giant resonance. Kiev: "Naukova Dumka", 1989 (in Ukrainian).

\bibitem{32}
L.I. Schiff, Phys. Rev. 83: 252 (1951), doi.org/10.1103/PhysRev.83.252.

\bibitem{33}
B.L. Zhuikov, M.V. Mebel, V.M. Kokhanyuk, et al., Phys. Rev. C 68: 054611 (2003), doi.org/10.1103/PhysRevC.68.054611.








\end{thebibliography}
\end{document}